# Determinants of Patent Citations in Biotechnology: An Analysis of Patent Influence Across the Industrial and Organizational Boundaries

**Antonio Messeni Petruzzelli**
(Corresponding author)
Department of Mechanics, Mathematics, and Management,
Politecnico di Bari
Viale Japigia 182 -70126 Bari, Italy
Tel. +39 080 5962776
Fax. +39 080 5982788
(antonio.messenipetruzzelli@poliba.it)

**Daniele Rotolo**
SPRU – Science and Technology Policy Research
University of Sussex
Brighton, BN1 9SL, United Kingdom

**Vito Albino**
Department of Mechanics, Mathematics, and Management
Politecnico di Bari
Viale Japigia 182 -70126 Bari, Italy

# Determinants of Patent Citations in Biotechnology: An Analysis of Patent Influence Across the Industrial and Organizational Boundaries

**ABSTRACT**

The present paper extends the literature investigating key drivers leading certain patents to exert a stronger influence on the subsequent technological developments (inventions) than other ones. We investigated six key determinants, as (i) the use of scientific knowledge, (ii) the breadth of the technological base, (iii) the existence of collaboration in patent development, (iv) the number of claims, (v) the scope, and (vi) the novelty, and how the effect of these determinants varies when patent influence—as measured by the number of forward citations the patent received— is distinguished as within and across the industrial and organizational boundaries. We conducted an empirical analysis on a sample of 5,671 patents granted to 293 US biotechnology firms from 1976 to 2003. Results reveal that the contribution of the determinants to patent influence differs across the domains that are identified by the industrial and organizational boundaries. Findings, for example, show that the use of scientific knowledge negatively affects patent influence outside the biotechnology industry, while it positively contributes to make a patent more relevant for the assignee's subsequent technological developments. In addition, the broader the scope of a patent the higher the number of citations the patent receives from subsequent non-biotechnology patents. This relationship is inverted U-shaped when considering the influence of a patent on inventions granted to other organizations than the patent's assignee. Finally, the novelty of a patent is inverted-U related with the influence the patent exerts on the subsequent inventions granted across the industrial and organizational boundaries.

# 1. INTRODUCTION

The growing importance of technological innovation for firms as a source of a sustainable competitive advantage [e.g. 1, 2] has led researchers to identify indicators to measure the outcomes of R&D process. Great attention has been specifically paid to patent data, which has been one of the most widely used sources of data among researchers for the evaluation of R&D outputs [3, 4]. A patent, as defined by the United States Patent and Trademark Office (USPTO), is "a property right granted by the Government of the United States of America to an inventor to exclude others from making, using, offering for sale, or selling the invention throughout the United States or importing the invention into the United States for a limited time in exchange for public disclosure of the invention when the patent is granted".[1] The temporary monopoly granted to a patent's assignee(s) (individuals, private or public organizations) generally lasts for 20 years. A patent is granted when it satisfies three main criteria: the invention the applicant aims to patent must (i) be novel, (ii) involve a non-obvious inventive step, and (iii) be capable of industrial application.[2] Several factors have contributed to the adoption of patent data for the aforementioned evaluation purposes [5]. First, patent data are available in most countries, especially in industrialized economies, where governments have collected data longitudinally. Second, patents contain a large amount of bibliographical information classified according to standardized schemes which in turn allows comparative analyses. Third, the extensiveness of data allows researchers and analysts to carry out both cross-sectional and longitudinal researches across multiple levels of analysis (e.g. nations, organizations, and individuals).

The potential of patent data to serve as an indicator of R&D outcomes has therefore attracted the attention of a large number of scholars. For instance, patent data have been proved capable to inform on different facets of the R&D process and outputs, such as the value of firms' intellectual property [6], efficiency of R&D processes [7] and technological position in the competitive landscape [8], as well as to assess the impact of innovation policies [9]. Significant research efforts have been particularly channeled in assessing the impact of patents and identifying those factors leading some patents to exert a stronger influence on subsequent technological developments (as mainly measured by forward citations counting) than other ones [e.g. 3, 10-12]. Despite these intensive research we have however limited knowledge on how the antecedents of patent influence differently affect the impact patents exerted on subsequent technologies when this impact is analyzed in different domains. In other words, it is less well understood whether patent influence is more or less bounded within specific domains. In this regard, the present paper aims at contributing to the extant literature on the determinants of patent influence by investigating how six key determinants previously identified in the literature (i) the use of scientific knowledge, (ii) the breadth of the technological base, (iii) the existence of collaboration in patent development, (iv) the number of claims, (v) the scope of the patent, and (vi) the novelty—differently shape the impact of patents in four domains as those identified within and across the industrial and organizational boundaries.

---

[1] The definition is reported in the USPTO Glossary available at www.uspto.gov/main/glossary. Other patent offices provide similar definitions.
[2] See the USPTO 2012 "Manual of Patent Examining Procedure" available at www.uspto.gov/web/offices/pac/mpep/.

We conducted our empirical analysis on a sample of 5671 (granted) patents registered at the USPTO by 293 US biotechnology firms from 1976 to 2003. The biotechnology industry is a suitable setting for our research given the importance of patents as an effective mean for the protection of intellectual property in this context [see also 13, 14]. Results reveal that the effect of the considered determinants varies according to the domain under investigation. For example, it emerges that the use of scientific knowledge negatively affects the influence a patent exerts on subsequent technologies outside the biotechnology industry, while it positively contributes to make the patent technologically more relevant for the assignee's future inventions; a patent of a broader scope impacts subsequent patents that are granted outside the biotechnology domains and to other firms, where its influence is positive and inverted U-shaped, respectively; the novelty of a patent has an inverted U-shape effect on patent influence across the industrial and organizational boundaries.

The remainder of the paper is organized as follows. Section 2 reviews the existing study investigating patent influence and describes the six main determinants we considered for our analysis. In Section 3, the research methodology is presented. Sections 4 and 5 report the results and discussion, respectively. Finally, Section 6 concludes the study.

## 2. THEORETICAL BACKGROUND
### 2.1. Measuring Patent Influence

The skewness characterizing the distribution the influence of patents on the subsequent technological developments makes the design of evaluation measures and tools a complex activity [15, 16]. Many attempts have been made in the extant literature to provide useful empirical strategies focused on information strongly correlated with patent influence. Some of the earlier research [e.g. 17] focused, for example, on the estimation of patent influence as reflected by the private value of patent rights, i.e. by analyzing the renewal data (e.g. patent renewals and renewal fee schedules). This approach has been subsequently validated and improved by several studies [7, 18]. The main logic underlying the adoption of renewal data is that the most valuable patents are those renewed to full statutory term, since renewing the patent protection for an additional year is expensive for the patent holders. However, this approach presents three main limitations [19]. First, for those patents renewed over their statutory lifetime, the renewal fees provide a lower rather than an upper bound for patent value. Second, this approach does not capture discontinuity in the patent value over its life until the patent reaches the maximum term. In fact, it is supposed that the annual returns from having the patent in force decrease monotonically over the patent life. Finally, renewal data provide information only on the part of patent value generally defined as patent premium [20], which represents the extra value that the invention generates to the assignee when it is patented.

We base our research on a complementary approach aimed at measuring the influence of patents on subsequent technological developments by the number of forward citations patents receive [e.g. 21, 22, 23]. The pioneering study by Trajtenberg demonstrated the existence of a positive and significant correlation between the social returns to innovation and forward citation indicators. Specifically, the author stated, "The cited patents opened the way to a technologically successful line of innovation [...]. Thus, if citations keep coming, it must be that the innovation originating in the cited patent had indeed proven to be valuable" [3: p.174]. Similarly, Albert et al. [24], focusing on the patents held by Eastman Kodak, showed the presence of a strong association between citation counts and the technical importance of patents. Patents' technological relevance, as reflected by forward citations, has been also proved to positively affect economic-based indicators, thus strengthening the strategic importance of assessing patents' technical performance. For example, Shane [25] found that, for a small sample of semiconductor firms, patents weighted by citations have more predictive power in a Tobin's Q equation than simple patent counts. Citations-weighted patents also turned out to be more highly correlated with R&D than simple patent counts. By analyzing a set of German patented inventions, Harhoff et al. [10] found that patents with greater economic relevance were more likely to be cited in subsequent patents. Giummo [26] found similar patterns by examining the royalties received by the inventor(s) and patent holder(s) at nine major German corporations under the German Employee Compensation Act. Lanjouw and Schankerman [27] also used citations, revealing that they have significant power to predict which patents will be renewed and which will be litigated. Hall et al. [11] revealed the existence of a positive relationship linking patent forward citations and the stock market valuation of firms' intangibles, showing that an extra citation per patent boosts market value by 3%. Finally, Gambardella et al. [19] employed data from an extensive European survey and found that patents' inventor economic return is significantly and positively correlated with the number of citations these receive.

In order to provide a better understanding of the role played by forward citations as proxy of patent influence, it is worth analyzing in more details what a patent document means. As discussed, a patent awards to one or more assignee (individuals, private or public organizations) the right to exclude others from the unauthorized use of the disclosed invention for a predetermined period of time. Patent citations define and limit the scope of the property rights by indicating the technological base upon which a specific patent is built [11]. Patent citations are added by the applicant, even if the decision regarding which citations to include ultimately rests with patent examiners, who may thus create noise in assessing the impact of patents, especially when citations are used as a proxy of knowledge flows [11, 28, 29]. Nevertheless, despite this limitation, forward citations remain one of the most suitable indicators to assess the influence of patents.

Apart from '*how*' measuring the influence of patents, a further question that has however received only scant attention in previous studies is '*where*' measuring the influence. A patented invention may influence the technological landscape of subsequent inventions within or across the specific industry in which it is granted. For example, the case of Viagra, originally developed by Pfizer for cardiovascular applications, has found significant application for the treatment of erectile dysfunction; a similar pattern emerged in the Corning's development of fiber optics, which have found important applications in the development of communication technologies [21]. On the other hand, the inventing firm or other organizations have exploited a patented technology to further develop technological solutions. The most classic example is represented by the 'EMI CAT' scanner created by the UK firm Electrical Musical Industries (EMI) Ltd. and then largely employed and refined by other companies that successfully dominated the market. Another case refers to Bowmar that invented the pocket calculator, which was however technologically exploited by Texas Instruments, Hewlett Packard and others [30]. Thereby, these cases highlight the importance to deepen our understanding of how patent influence varies across domains—identified by the industrial and organizational boundaries—to properly assess the impact of patented technological solutions as well as to design R&D strategies and policies aimed at maximizing the output of the inventing process.

## 2.2. Antecedents of Patent Influence

Drawing from previous literature, we focus the attention on six determinants of the influence of patents: (i) the use of scientific knowledge, (ii) the breadth of the technological base, (iii) the existence of collaboration in patent development, (iv) the number of claims, (v) the scope of the patent, and (vi) the novelty.

***Scientific Knowledge:*** Scholars have discussed the significant contribution of science to economic growth for years [31, 32]. Research has demonstrated that knowledge flowing from public research organizations makes a considerable contribution to industrial innovations and, consequently, to welfare. For instance, Mansfield [33], analyzing 76 US firms in seven industries, found that 11% of product innovations and 9% of process innovations would not have been developed in the absence of recent academic research. Fleming and Sorenson [34] suggested that science functions as a map of the technological landscapes that can guide private research towards the most economically promising technological areas, thus avoiding wasteful experimentations. Nevertheless, regarding the relationship between the use of scientific knowledge (non-patent references) and patent impact, contrasting results can be found [35]. While Henderson et al. [36] and Mowery et al. [37] found that academic patents receive more citations than non-university patents, thus confirming the importance of scientific knowledge as input in the inventing process, other authors have provided divergent results, showing that links to science exert different effects on the impact of resulting patents according to the firm's industry, the inventive problems to be addressed, and the level of analysis (firm or patent) [34, 38, 39].

***Breadth of Technological Base:*** Scholars have emphasized that innovativeness often depends on firms' ability to search for knowledge widely [e.g. 23, 39], since a broader search increases the likelihood of discovering new and useful knowledge combinations [40]. However, the search process is highly path-dependent and constrained by organizational routines [41], hence leading organizations to search for knowledge locally, i.e., into the neighbor of their existing capabilities [42]. The relationship between searching widely and patent influence has been extensively discussed. For example, by analyzing the innovation processes in the optical disk industry, Rosenkopf and Nerkar [23] found that patented innovations reflecting broader search efforts also have a greater impact on subsequent technologies. Using data on over half a million patents, Singh [43] revealed that patents based on a broad range of technologies present a higher quality. Finally, Capaldo and Messeni Petruzzelli [44] showed the positive impact exerted by the number of different knowledge domains across which firms search on the number of forward citations received by their patents. A broader search involves new interaction between the components, thus constituting the basis for a broader range of subsequent innovative developments and contributing to the pace of the technological progress [45]. A broader search process may also expand the patent life, then increasing the monopolistic rents of the holder organization [46]. However, increasing the breadth of technological search may have harmful consequences on the development of successful inventions. In fact, searching widely is a risky process that may lead to uncertain and unpredictable outcomes. Furthermore, searching across and combining a great variety of technological components may create several difficulties due to a lack of absorptive capacity, common knowledge base, and inexperience or unfamiliarity with the recombined technologies [47].

***Joint Development:*** Developing innovation can be conceived as a process where complementary and heterogeneous inputs (i.e., pieces of knowledge) are transformed into outputs (i.e., results of innovations) [48]. The increasing complexity of the knowledge-creation processes has lead firms to rely upon external resources [49] and hence search beyond their own organizational boundaries for valuable knowledge and skills complementing their capabilities [50]. In fact, innovation partly depends on firm-specific knowledge resources and strongly depends on determinants external to the firm. This is because firms often specialize in one or few fields of knowledge and rarely have all the required resources internally. Therefore, to innovate successfully, firms need to collaborate to gain access to knowledge resources that are not internally available [51]. Collaborations allow organizations to expand their knowledge base and, thus, to explore new opportunities and solutions [52], which in turn may lead to the development of technologically valuable innovations [e.g. 53]. Nevertheless, despite these advantages, collaborations may also hamper the result of the innovation process due to the emergence of opportunistic behaviors and differences between partners' bargaining power [e.g. 54]. This may increase competition and reduce the benefits going along with inter-organizational resource integration. Inter-organization collaborations at the patent level may be captured by analyzing the presence of co-assignees, i.e. whether patent property rights are jointly shared between two or more organizations [55, 56]. Co-owned patents may assume an important role in industries with strong regimes of appropriability, such as chemical, pharmaceutical, and biotechnology [55]. In addition, joint patents are largely the result of inter-organizational collaborations where the companies are unable to 'divide' the invention among the partners, hence creating the necessity to share the intellectual property [57].

***Claims:*** The claims reported in patent documents define the legally enforceable aspects of the given invention. Patent claims can be distinguished as principal and subordinate. The former defines the essential novel features of the invention, whereas the latter describes detailed features of the innovation. The number of claims can affect patent influence: the broader the property rights protection the higher the probability that others will rely upon the invention. This explains the incentive leading assignees in increasing the number of claims in the patent application [58]. Tong and Frame [59] were the first to use information on claims to conduct empirical analysis. Specifically, they investigated the relationship between the number of claims reported in patents and several macroeconomic (nation-level) indicators of technological development. Later, Lanjouw and Schankerman [27] adopted the information on claims to test their influence on the likelihood of challenge and validity suits for a sample of US patents. Recently, Bonaccorsi and Thoma [60] employed the number of patent claims as an indicator to analyze how the quality of patents varies across different communities of inventors.

***Scope:*** One of the most discussed determinants of patent impact is represented by its technological scope. The scope of a patent may be an important determinant of the efficacy of patent protection [61]. In their pioneering study, Merges and Nelson evidenced how "[...] the broader the scope, the larger the number of competing products and processes that will infringe the patent [...]" [62: p. 839], thus increasing the technological influence of the patent. The relevance of the scope to enhance patent influence has been also investigated by several empirical works. For instance, Lerner [63] demonstrated that the patent scope has a significant and positive effect on the valuation of venture capitalists financing biotechnology start-ups. Shane [64], conducting an extensive analysis of 1,397 patents assigned to the Massachusetts Institute of Technology, showed that the scope of patents increases the likelihood of the relative inventions to be commercialized.

***Novelty:*** One of the central concepts in the innovation study is that refining and improving an existing technology and introducing a new approach to technical practices are fundamentally different things [65]. The development of novel technologies is generally associated with the first stage of the technology life-cycle, being thus characterized by market uncertainty and R&D efforts [66]. However, if successful, technologies coming through this phase and moving to the growth stage have more opportunities to become radical innovations, hence breaking existing technological paradigms [64], shifting towards new trajectories, and representing the basis on which further innovations are built. Therefore, novel patented innovations may represent rare, valuable, and inimitable sources of competitive advantage for firms [67], allowing business growth and new business development [13]. Such an advantage mainly derives from the benefits associated with learning economies [68], causal ambiguity [69], switching costs [70], consumer learning, and reputation advantages [71]. Nevertheless, as discussed, developing a novel patented innovation is an uncertain and risky task [72]. Environmental factors, such as the pace of technology and market evolution [73], may also affect the relevance of such novel technical solutions. In particular, technological uncertainty makes buyers reluctant to invest in product-specific competencies [70, 71, 74], with the consequent risk of introducing an underdeveloped product that will make final customers more willing to switch to alternative products [75]. In addition, market evolution may imply changes in consumer tastes or preferences, emergence of new regulations, degree of market fragmentation, and consumer learning [76], which may in turn impact the diffusion and success of a novel patented innovation.

## 3. METHODS

### 3.1. Research Setting and Sample Data

We conducted out empirical analysis on the US biotechnology industry. The rise of biotechnology has its origins in the discovery of the double helix structure of DNA by Watson and Crick in 1953 and the subsequent development of DNA recombination by Cohen and Boyer in 1973. The latter can be considered a radical innovation process that broke the barriers of entry into the pharmaceutical industry [77, 78]. We chose this particular research setting since the biotechnology industry is one of the most innovation-intensive industries [79], and the patenting activity within this context plays a critical strategic role for firms' performance [80-82]. We focus our attention on the US market representing the most significant center of innovation for biotechnology [14, 80, 83]. For instance, in 2006, the US market was revealed as the market with the highest concentration of dedicated biotechnology firms, which spent US$25,101 million on R&D, accounting for 75% of total biotechnology R&D expenditures in developed countries [84].

One of the major criticisms advanced against the adoption of patent data is represented by the unobserved heterogeneity across industries and technology fields [85], which can significantly affect the suitability of patents as indicators of R&D outcomes. However, in the biotechnology field, innovations are more discrete in terms of product fragmentation, and, hence, may be covered by a small number of patents [80]. Individual patents can yield substantial rents from commercialization and/or licensing. For this reason, firms operating in this industry are likely to dedicate significant resources on the writing of strong and relevant patent applications. This supports the use of patent data as a proxy of the R&D outcomes.

For the empirical analysis, we relied on a patent dataset drawn from a population of 358 US public and private firms included in the BioScan database in 2010. BioScan is one of the most recognized biotechnology industry reporting services [13]. We searched for patents firms filed under the US patent system during the observation period 1976–2003 by querying the USPTO database.[3] We included in the sample only those firms that granted at least one biotechnology patent in the observation period—we identified the industrial boundary by using the US patent technological classes [86].[4] This reduced the sample of firms from 358 to 293. Those firms granted 5671 patents, which represent our final sample and unit of analysis.[5] For each patent, we collected bibliographical data and backward and forward patent citations. Analyzing distribution of patents across the firms in our sample (see Figure 1), Genentech, Pioneer Hi-Bred, and Chiron emerge as the main players in terms number of patents (14.44%, 12.03%, and 10.67%, respectively).

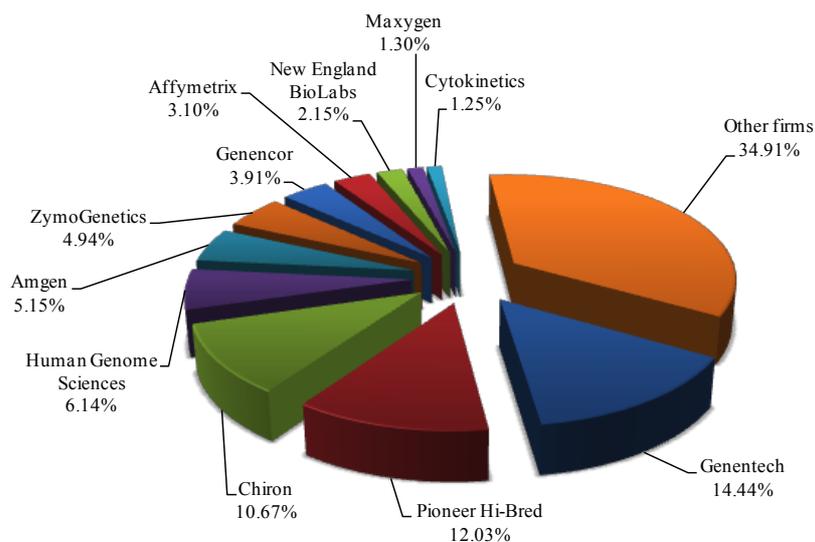

Figure 1. Distribution of biotechnology patents across the firms in the sample.

---

[3] We selected the 1976–2003 period for the following reasons. First, the conceptualization and development of the key operational principles of biotechnology paradigm, that are the genetic engineering and monoclonal antibodies, started in early 1970s and first venture capitalists recognized the utility of these techniques in the late 1970s [78]—for example, Genentech, the first new biotechnology firm, was founded in 1976. Second, the USPTO allows electronic search for patents starting from 1976. Finally, since the data were collected in 2009, we ended the observation period in 2003 in order to allow the patents to be cited in an equal five-year time window.

[4] Building on previous studies [86], the boundaries of the biotechnology domain were defined using the following three-digit US technological classes: 424, 435, 436, 514, 530, 536, 800, 930, and PLT.

[5] The number of patents grown rapidly from 1976 up to 2003 with a peak of 981 and 739 patents in 1995 and 2000, respectively.

### 3.2. Variables

***Dependent Variables:*** As discussed in the previous section, we estimated the influence of a patent on subsequent technologies as the number of forward citations the patent received [3, 4, 15, 60, 87]. The use of forward citations is particularly suitable for our study, since citations added by examiners do not represent a critical issue for two main reasons. First, we analyze the biotechnology context, where the share of applicants attributed forward citations is on average greater than 70% [28] and assignees strategically withhold only 5–7% of forward citations [29]. Second, we rely on patents granted by the USPTO to US firms, hence further reducing the share of examiner citations, which has been proven to be especially high among foreign firms [28].[6] To control for the fact that older patents have a higher likelihood of being cited by following patents, we considered only those citing patents filed within a five-year time window after the given patent's filing date [42].

With the aim to disentangle the influence of patents across different domains, we distinguished it according to the industrial (biotechnology) and organizational boundaries (see Figure 2). Firstly, we considered the whole patent influence (*Influence*), which was measured by the number of forward citations a patent received from subsequent patents. Following previous research [3, 87], the whole influence of the patent *i* (*1,2,…,m*) is assessed as follows:

$$Influence_{i,t} = 1 + \left( \sum_{t}^{t+4} Cit_{i,t} \right),$$

where *t* (*1976,…,2003*) refers to the *i-th* patent's filing year and $Cit_{i,t}$ is the number of forward citations the *i-th* patent received from following patents up to five years after the filing date. Secondly, we classified patent influence as within (*InfluenceInBio*) and outside (*InfluenceOutBio*) the boundary of the biotechnology industry [4, 23]. In particular, a patent citing the given patent *i* is classified as inside the biotechnology industry if it is assigned to at least one of the nine US technology classes defining the boundary of the biotechnology industry (see footnote 4), otherwise it is classified as outside the biotechnology. Thus, a patent's influence within and outside the industrial boundary is evaluated as:

$$InfluenceInBio_{i,t} = 1 + \left( \sum_{t}^{t+4} CitInBio_{i,t} \right)$$

$$InfluenceOutBio_{i,t} = 1 + \left( \sum_{t}^{t+4} CitOutBio_{i,t} \right)$$

---

[6] The share of examiners' citations in our sample of patents, which may be accounted for only the 410 patents granted after 2000, is about 22% of the overall number of forward citations.

where $t$ (*1976,…,2003*) refers to the *i-th* patent's filing year and *CitInBio*$_{i,t}$ and *CitOutBio*$_{i,t}$ represent the number of forward citations the *i-th* patent received by the subsequent patents assigned inside and outside the biotechnology industry within a 5-year time window, respectively. Finally, referring to the organizational boundary, the influence of a patent is distinguished into *InfluenceInFirm* and *InfluenceOutFirm* if the citing patent is granted to the same assignee as the given patent or to a different one [4, 88]. Therefore, we adopted the following measures:

$$InfluenceInFirm_{i,t} = 1 + \left( \sum_{t}^{t+4} CitInFirm_{i,t} \right)$$

$$InfluenceOutFirm_{i,t} = 1 + \left( \sum_{t}^{t+4} CitOutFirm_{i,t} \right)$$

where $t$ (*1976,…,2003*) refers to the *i-th* patent's filing year and *CitInFirm*$_{i,t}$ and *CitOutFirm*$_{i,t}$ represent the number of forward citations the *i-th* patent received by the subsequent patents granted to the same firm assignee or other assignees within a 5-year time window, respectively.

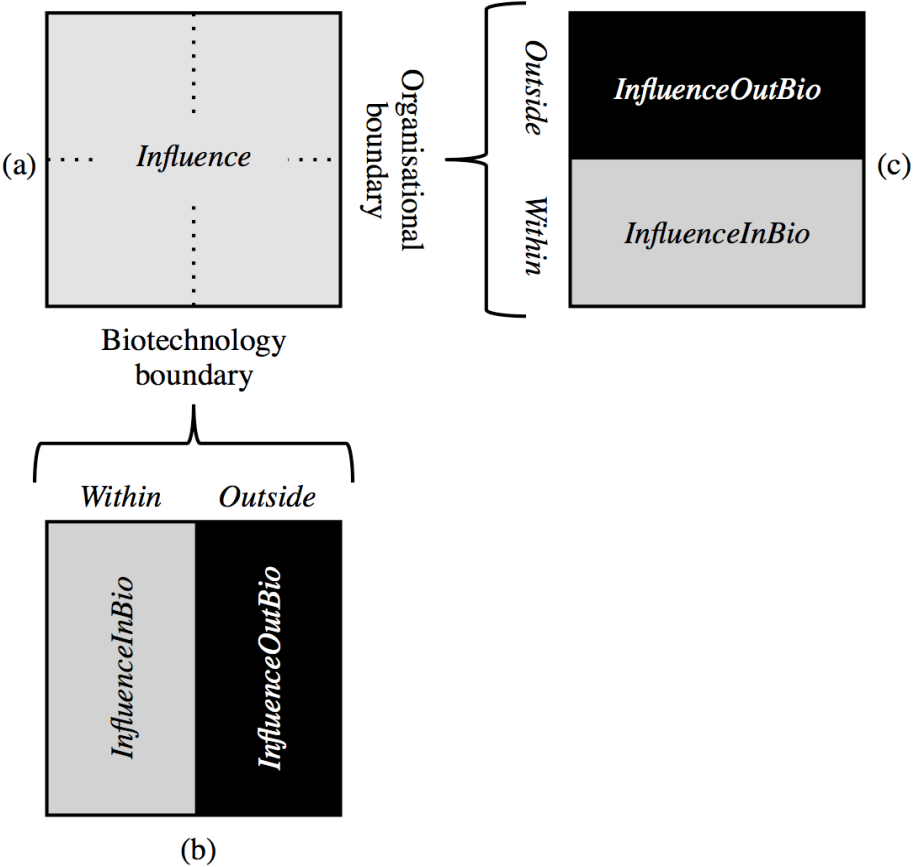

Figure 2. Patent influence across the industrial (biotechnology) and organizational boundaries.

To validate our patent influence indicators, we conducted a robustness test by analyzing the distribution of the family size of patents, which has been proven to be an alternative predictor of the relevance underlying an invention [27, 89]. Specifically, we referred to family size as "the number of different patent systems in which protection for a single invention is sought" [89: p. 2]. Thus, even though family size is not suitable for the purpose of the present paper, since it does not allow patent influence to be distinguished across multiple domains, we compared our citation-based measure with patent family size. We found our measures of patent influence strongly related to family size, thus bestowing further confidence on our choice.

***Independent Variables:*** To account for the use of scientific knowledge (*ScienKnowledge*) in the inventing process, we measured the number of scientific non-patent references the given patent cited. The number of non-patent backward citations is a suitable measure to evaluate the extent to which a patent is based on scientific knowledge [10, 90].[7] To measure the breadth of the technological base (*TechBreadth*) upon which the given patent is built, we relied on the measure proposed by Jaffe and Trajtenberg [4]. The construction of this measure follows three steps: (i) identification of all citations made by the given patent *i*, (ii) identification of the technological classes assigned to the cited patents and (iii) computation of the index equal to one minus the Herfindahl concentration index. Thus, it results in the following equation:

$$TechBreadth = 1 - \sum_{j}^{n} s_{ij}^2,$$

where $s_{ij}$ refers to the fraction of patents cited by patent *i* that belong to technological class *j* out of *n* technological categories assigned to the patents by the USPTO. Specifically, we refer to the three-digit classes [91]. Since this variable cannot be defined when there are no backward citations, in this case, technological breadth is set equal to zero. The joint development of patents (*JointDevelopment*) was computed by counting the number of co-applicants to which the patent was granted [35, 55].[8] To account for the effect exerted by claims, we measured the number of claims per patent (*Claims*) [19, 27, 59]. Following the approach proposed by Lerner [63], we measured the scope of a patent (*Scope*) as a time-

---

[7] Results are confirmed when we measured the use of scientific knowledge as a dummy variable taking value one if the focal patent cites at least one non-patent reference, and zero otherwise.
[8] Inter-organizational collaborations were measured also as a dummy variable taking value one if the focal patent is assigned to more than one firm, and zero otherwise. The results did not significantly changed.

-14-

invariant count of the number of three-digit technological classes the USPTO assigned to the patent [10]. Finally, the novelty of a patent (*Novelty*) was evaluated as the number of three-digit technological classes in which patents cited by the patent are found, but the patent itself is not classified [64].

Table 1. Definitions of the dependent, independent, and control variables.

| Variable | Description |
| --- | --- |
| *Dependent variables* | |
| Influence | Number of citations a patent received |
| InfluenceInBio | Number of citations a patent received from patents within the biotechnology domain |
| InfluenceOutBio | Number of citations a given patent received from patents outside the biotechnology domain |
| InfluenceInFirm | Number of citations a patent received from patents granted to the same assignee |
| InfluenceOutFirm | Number of citations a patent received from patents granted to a different assignee |
| | |
| *Independent variables* | |
| ScienKnowledge | Number of non-patent references a patent cited |
| TechBreadth | Breath of the technological base upon which a patent is built |
| JointDevelopment | Number of patent co-applicants |
| Claims | Number of patent claims |
| Scope | Number of US three-digit technological classes assigned to a patent |
| Novelty | Number of US three-digit technological classes in which patents cited by a patent are found, but wherein the patent itself is not classified |
| | |
| *Control variables* | |
| FirmPatents | Number of patents the assignee filed at the USPTO during a five-year time window preceding the filing date of the given patent |
| FirmAge | Number of years elapsed between the assignee's year of incorporation and the filing year of the given patent |
| TeamSize | Number of inventors involved in a patent |
| BackCitations | Number of patent backward citations |
| CitAge | Coefficient of variation (standard deviation/mean) of the number of years elapsed from the filing date of the cited patents to the filing date of a patent |
| GovInt | Dummy variable that takes value one if a patent was funded by the U.S. government, zero otherwise |
| SIC Code | Dummy variable indicating the main assignee's SIC code |
| Year | Dummy variables indicating a particular year in the observation period (1976–2003 period) |

Note: *Influence*, *InfluenceInBio*, *InfluenceOutBio*, *InfluenceInFirm*, and *InfluenceOutFirm* variables count the number of citations a patent received within a five-year time window.

***Control Variables:*** We also included several control variables that may explain patent influence. First, we controlled for firms' patent stock (*FirmPatents*), which is used to take into account the technological capital owned by the biotechnology firm to which the given patent is granted [13, 14, 92]. The firm's patent stock was measured as the number of patents the firm granted at the USPTO during a 5-year time window preceding the filing date of the given patent. Second, we controlled for the firm's age (*FirmAge*) as the number of years elapsed between the firm's year of incorporation and the patent's filing year [79]. Third, the size of the team involved in a patent's development can have a direct effect on its relevance, because of economies of specialization, a larger and more diverse pool of knowledge, and access to a wider and more heterogeneous external network. Therefore, we included the number of inventors involved in the given patent (*TeamSize*) [43]. Fourth, we controlled for the patent's number of backward citations (*BackCitations*) [10]. Fifth we controlled for the age of the patents the focal patent cited, i.e. the number of years elapsed from the filing date of the cited patents to the filing date of the focal patent. We specifically included a variable (*CitAge*) measuring the coefficient of variation (standard deviation/average) of the age of cited patens.[9] Sixth, we controlled whether a patent was financially supported by the US government (*GovInterest*). This was evaluated by a dummy variable that takes a value one if the given patent was funded by US governmental agencies and zero otherwise. Finally, we also included dummy variables (33) for the main SIC codes assigned to the patents' assignees and year dummies (26). Table 1 summarizes our variables.

## 3.3. Estimation Procedure

The dependent variables are count variables, taking only integer and positive values. Thus, the use of linear regression modeling is inadequate since the distribution of residuals will be heteroskedastic non-normal. In this case, the use of a Poisson regression approach is preferred. This estimation however assumes the mean and variance to be equal while patent data often present over-dispersion, i.e. the variance exceeds the mean [93]. The over-dispersion is confirmed by the coefficients of variation equal to 2.30, 2.66, 4.22, 2.77, and 2.82 for *Influence*, *InfluenceInBio*, *InfluenceOutBio*, *InfluenceInFirm*, and *InfluenceOutFirm*, respectively. Therefore, as for previous studies [e.g. 23], we found the negative binomial estimation as more suitable to our data—this specification can handle over-dispersion since it allows the variance to differ from the mean [93]. The negative binomial estimation considers a variable $y_i$ that follows a Poisson regression model with parameter $\lambda_i$ and omitted variable $u_i$ such that $exp(u_i)$ follows a gamma distribution with mean one and variance $\alpha$:

---

[9] By introducing the coefficient of variation of the age of cited patents we further reduced the risk of multicollinearity issues in the regression models since the average and standard deviation tend to be correlated—in our data the correlation between the average and standard deviation of the age of cited patents is equal to 0.530.

$$y_i \sim \text{Poisson}(\mu_i^*),$$

$$\mu_i^* = \exp(x_i \varphi_i + u_i), \text{ and}$$

$$\exp(u_i) \sim \text{Gamma}(1/\alpha, 1/\alpha)$$

The $\varphi_i$ is the vector of the parameter associated with the vector of independent variables $x_i$ and $\alpha$ is the over-dispersion parameter. We reported the model specification in the following:

$$\hat{\lambda}_i = \exp\left(\sum_m \beta_m IndepVar_i + \sum_k \delta_k ContrVar_i + \varepsilon_i\right)$$

where $i=1, ..., 5,671$, $\varepsilon_i$ is the error term, and $[\beta_1, ..., \beta_m, \delta_1, ..., \delta_k]$ is vector of the parameter to estimate—$\beta_m$ refers to the independent variables (*IndepVar*) and $\delta_k$ refers to the control variables (*ContrVar*). The models were estimated through the "nbreg" routine in the STATA 10.0 software package. Significance levels are based on Huber-White robust standard errors.

## 4. RESULTS

Table 2 report the descriptive statistics and correlation between our variables. Correlations among the independent variables are low such that multicollinearity can be considered to be not a cause for concern. We also standardized the independent variables before creating the squared terms and running the regression. The standardization reduces potential multicollinearity issues in the estimation process when testing the curvilinear effect of an independent variable [94].

In Table 3, we reported the results of the negative binomial estimations of patent influence across the different domains identified by the industrial and organizational boundaries: Model 1 report the analysis for patent *Influence*, whereas Models 2–5 test patent influence as distinguished in *InfluenceInBio*, *InfluenceOutBio*, *InfluenceInFirm*, and *InfluenceOutFirm*, respectively. We reported the estimation of the over-dispersion parameter alpha ($\alpha$) for each model. The results confirm the negative binomial estimation to be more suitable than the Poisson one since the estimated over-dispersion parameter is significantly different from zero in each model. In addition, when the squared term of an independent variable was found not statistically significant in the regression model, we ran another regression excluding this term. This allowed us to obtain better estimates by further reducing the risk of multicollinearity issues.

Model 1 provides evidence that the number of claims (*Claims*, $\beta = 0.1896$, $p < 0.001$; *Claims²*, $\beta = -0.0061$, $p < 0.001$) and the novelty of a patent (*Novelty*, $\beta = 0.2658$, $p < 0.001$; *Novelty²*, $\beta = -0.0423$, $p < 0.1$) are curvilinearly (inverted-U) related with the patent's influence. The results are different when analyzing the influence within the biotechnology domain (Model 2). Specifically, while the inverted U-shape effect of the number of claims (*Claims*, $\beta = 0.1866$, $p < 0.001$; *Claims$^2$*, $\beta = -0.0059$, $p < 0.001$) is in line with the relationships we found for *Influence* (Model 1), the involvement of more than one firm in the development of a patent (*JointDevelopment, $\beta$=0.0960, $p < 0.1$*) also exerts a positive effect on the number of citations the patent receives from subsequent domain-specific (biotechnology) patented inventions. Model 3 reveals that both *ScienKnowledge* ($\beta = -0.0022$, $p < 0.001$) and *TechBreadth* ($\beta = -0.0419$, $p < 0.1$) negatively affect the influence of a patent on subsequent patents granted outside the industrial boundaries (*ValueOutBio*). Differently, *Claims* ($\beta = 0.0645$, $p < 0.01$) and *Scope* ($\beta = 0.0762$, $p < 0.001$) show a positive impact on *InfluenceOutBio*. Finally, the novelty (*Novelty*, $\beta = 0.4590$, $p < 0.001$; *Novelty$^2$*, $\beta = -0.0467$, $p < 0.001$) curvilinearly affects (inverted-U) the influence the patent exerts on subsequent invention outside the industry domain. In Model 4, we tested the determinants of patent influence within the assignee's organizational boundaries. The use of scientific knowledge (*ScienKnowledge*, $\beta = 0.0014$, $p < 0.01$) is positively related with *InfluenceInFirm*, whereas the number of claims (*Claims*, $\beta = 0.1650$, $p < 0.001$; *Claims$^2$*, $\beta = -0.0045$, $p < 0.001$) confirms the inverted-U shaped effect supported in Model 1 and Model 2. In Model 5, the effect of the antecedents on the impact of patents across the assignee's organizational boundaries is tested. The results show the number of claims (*Claims*, $\beta = 0.1648$, $p < 0.001$; *Claims$^2$*, $\beta = -0.0061$, $p < 0.001$), the scope (*Scope*, $\beta = 0.0517$, $p < 0.1$; *Scope$^2$*, $\beta = -0.0174$, $p < 0.05$), and the novelty measure (*Novelty*, $\beta = 0.2905$, $p < 0.001$; *Novelty$^2$*, $\beta = -0.0413$, $p < 0.001$) exerting a non-monotonic impact (inverted-U) on the number of citations the patent received from companies other than that of the assignee. Finally, referring to the control variables, it is worth noting the positive effect of the number of inventors (*TeamSize*) and the negative impact of the coefficient of variation of the age of the cited patents (*CitAge*) on patent impact in the different domains, except for patent influence across the industrial field (Model 3) and within the organization context (Model 4).

## 5. DISCUSSION

Overall, we revealed that the six key determinants we analyzed differently contribute to explain the influence of patents as this is decomposed according to the industrial and organizational boundaries. Table 4 summarizes the main results. Figure 3 reports the slope analysis and plots the relationships between the six determinants and patent influence.

Table 2. Descriptive statistics and correlation matrix (N = 5671).

| Variable | 1 | 2 | 3 | 4 | 5 | 6 | 7 | 8 | 9 | 10 | 11 | 12 | 13 | 14 | 15 | 16 | 17 |
|---|---|---|---|---|---|---|---|---|---|---|---|---|---|---|---|---|---|
| 1. Influence | 1.000 | | | | | | | | | | | | | | | | |
| 2. InfluenceInBio | 0.974 | 1.000 | | | | | | | | | | | | | | | |
| 3. InfluenceOutBio | 0.508 | 0.301 | 1.000 | | | | | | | | | | | | | | |
| 4. InfluenceInFirm | 0.570 | 0.562 | 0.263 | 1.000 | | | | | | | | | | | | | |
| 5. InfluenceOutFirm | 0.980 | 0.953 | 0.505 | 0.393 | 1.000 | | | | | | | | | | | | |
| 6. ScienKnowledge | 0.022 | 0.025 | 0.000 | 0.037 | 0.016 | 1.000 | | | | | | | | | | | |
| 7. TechBreadth | 0.074 | 0.055 | 0.104 | 0.027 | 0.076 | 0.224 | 1.000 | | | | | | | | | | |
| 8. JointDevelopment | -0.004 | 0.002 | -0.026 | -0.017 | -0.001 | 0.043 | 0.045 | 1.000 | | | | | | | | | |
| 9. Claims | 0.024 | 0.015 | 0.043 | 0.046 | 0.015 | 0.106 | 0.041 | 0.011 | 1.000 | | | | | | | | |
| 10. Scope | 0.079 | 0.056 | 0.120 | 0.025 | 0.082 | 0.054 | 0.147 | 0.028 | 0.062 | 1.000 | | | | | | | |
| 11. Novelty | 0.090 | 0.043 | 0.221 | 0.041 | 0.092 | 0.427 | 0.429 | 0.007 | 0.088 | 0.072 | 1.000 | | | | | | |
| 12. FirmPatents | -0.117 | -0.097 | -0.121 | -0.047 | -0.119 | 0.072 | 0.012 | -0.015 | 0.002 | 0.044 | -0.149 | 1.000 | | | | | |
| 13. FirmAge | -0.068 | -0.048 | -0.104 | 0.053 | -0.089 | -0.089 | -0.099 | -0.064 | -0.077 | 0.003 | -0.174 | 0.201 | 1.000 | | | | |
| 14. TeamSize | -0.001 | -0.005 | 0.018 | -0.011 | 0.002 | 0.080 | 0.075 | 0.221 | 0.118 | 0.063 | 0.073 | 0.010 | -0.140 | 1.000 | | | |
| 15. CitAge | 0.030 | 0.017 | 0.064 | 0.022 | 0.029 | 0.104 | 0.430 | 0.035 | 0.072 | 0.051 | 0.277 | -0.121 | -0.110 | 0.071 | 1.000 | | |
| 16. BackCitations | 0.053 | 0.023 | 0.136 | 0.036 | 0.051 | 0.534 | 0.289 | -0.012 | 0.128 | 0.090 | 0.572 | -0.103 | -0.134 | 0.069 | 0.224 | 1.000 | |
| 17. GovInterest | 0.040 | 0.022 | 0.083 | 0.032 | 0.037 | 0.015 | 0.026 | 0.129 | 0.010 | 0.012 | 0.026 | -0.071 | -0.073 | 0.092 | 0.013 | 0.007 | 1.000 |
| Mean | 10.25 | 9.80 | 2.45 | 2.03 | 8.22 | 31.23 | 0.532 | 0.099 | 20.09 | 2.273 | 4.062 | 109.8 | 20.11 | 3.004 | 0.351 | 9.138 | 0.033 |
| Std. Dev. | 25.93 | 23.42 | 6.13 | 5.63 | 23.16 | 48.53 | 0.293 | 0.315 | 21.73 | 0.998 | 6.553 | 131.1 | 22.20 | 2.021 | 0.383 | 19.61 | 0.180 |
| Min | 1 | 1 | 1 | 1 | 1 | 0 | 0 | 0 | 0 | 1 | 0 | 0 | 0 | 1 | 0 | 0 | 0 |
| Max | 305 | 296 | 62 | 71 | 266 | 438 | 0.931 | 2 | 683 | 8 | 63 | 531 | 137 | 27 | 11.56 | 259 | 1 |

Table 3. Negative binomial regression on patent influence across different domains (N = 5671).

| | Model 1 (*Influence*) | | Model 2 (*InfluenceInBio*) | | Model 3 (*InfluenceOutBio*) | | Model 4 (*InfluenceInFirm*) | | Model 5 (*InfluenceOutFirm*) | |
|---|---|---|---|---|---|---|---|---|---|---|
| *Independent variables* | | | | | | | | | | |
| ScienKnowledge | 0.0004 | (0.0005) | 0.0008 | (0.0005) | -0.0022*** | (0.0005) | 0.0014** | (0.0005) | 0.0002 | (0.0005) |
| TechBreadth | 0.0018 | (0.0331) | 0.0400 | (0.0322) | -0.0419† | (0.0215) | 0.0280 | (0.0240) | -0.0147 | (0.0417) |
| JointDevelopment | 0.0803 | (0.0539) | 0.0960† | (0.0545) | 0.0143 | (0.0528) | -0.0106 | (0.0589) | 0.0814 | (0.0540) |
| Claims | 0.1896*** | (0.0236) | 0.1866*** | (0.0236) | 0.0645** | (0.0215) | 0.1650*** | (0.0237) | 0.1648*** | (0.0243) |
| Claims$^2$ | -0.0061*** | (0.0010) | -0.0059*** | (0.0010) | | | -0.0045*** | (0.0010) | -0.0061*** | (0.0011) |
| Scope | 0.0153 | (0.0206) | -0.0051 | (0.0213) | 0.0762*** | (0.0170) | -0.0099 | (0.0199) | 0.0517† | (0.0289) |
| Scope$^2$ | | | | | | | | | -0.0174* | (0.0088) |
| Novelty | 0.2658*** | (0.0486) | 0.0461 | (0.0488) | 0.4590*** | (0.0471) | 0.0690 | (0.0547) | 0.2905*** | (0.0520) |
| Novelty$^2$ | -0.0423*** | (0.0062) | | | -0.0467*** | (0.0062) | | | -0.0413*** | (0.0064) |
| *Control variables* | | | | | | | | | | |
| FirmPatents | -0.0211 | (0.0260) | -0.0172 | (0.0266) | -0.0044 | (0.0215) | -0.0904*** | (0.0230) | -0.0157 | (0.0288) |
| FirmAge | -0.0013 | (0.0033) | -0.0016 | (0.0033) | -0.0076** | (0.0028) | -0.0043 | (0.0029) | 0.0008 | (0.0035) |
| TeamSize | 0.0625*** | (0.0169) | 0.0674*** | (0.0170) | 0.0089 | (0.0184) | 0.0307 | (0.0199) | 0.0680*** | (0.0184) |
| CitAge | -0.1416** | (0.0457) | -0.1450** | (0.0455) | -0.0589 | (0.0450) | 0.0572 | (0.0628) | -0.1907*** | (0.0544) |
| BackCitations | 0.1030* | (0.0468) | 0.1917*** | (0.0579) | 0.0352 | (0.0462) | -0.0316 | (0.0526) | 0.0855* | (0.0431) |
| BackCitations$^2$ | | | -0.0211*** | (0.0037) | | | | | | |
| GovInterest | -0.0431 | (0.0954) | -0.0087 | (0.0986) | 0.0862 | (0.0983) | 0.1699 | (0.1102) | -0.0973 | (0.0961) |
| SIC Code | Included | | Included | | Included | | Included | | Included | |
| Year | Included | | Included | | Included | | Included | | Included | |
| *Intercept* | 2.9711*** | (0.3750) | 2.2830*** | (0.3791) | 3.1207*** | (0.3275) | 1.0649** | (0.3272) | 2.7858*** | (0.4049) |
| Log pseudo-likelihood | -18122.71 | | -17527.43 | | -10090.10 | | -12213.85 | | -16793.01 | |
| Degree of freedom | 63 | | 63 | | 62 | | 62 | | 64 | |
| Alpha (over-dispersion) | 0.916*** | (0.030) | 0.912*** | (0.031) | 0.397*** | (0.021) | 0.647*** | (0.025) | 0.908*** | (0.042) |

Notes: †*p* < 0.1; **p* < 0.05; ***p* < 0.01; ****p* < 0.001;
Huber-White robust standard errors in parentheses;
The hypothesis test for the over-dispersion is that the alpha parameter is zero.

We specifically found that the use of scientific knowledge negatively affects a patent's impact on subsequent patents that are granted outside the biotechnology industry, while it increases the technological relevance of the patent for the assignee's future inventions. A more extensive use of scientific knowledge may provide assignees with a broader knowledge base to source for subsequent inventions [34, 95]. On the other hand, the negative effect of the use of scientific knowledge on the number of citations a patent receives from subsequent non-biotechnology patents may be due to the greater complexity that features patents building on a broader scientific knowledge base [39]. This in turn contributes to reduce the degree of understanding and applicability in sectors other than biotechnology.

Table 4. Determinants and main effects on patent influence.

| | *Influence* | *InfluenceInBio* | *InfluenceOutBio* | *InfluenceInFirm* | *InfluenceOutFirm* |
|---|---|---|---|---|---|
| *ScienKnowledge* | n.s. | n.s. | - | + | n.s. |
| *TechBreadth* | n.s. | n.s. | - | n.s. | n.s. |
| *JointDevelopment* | n.s. | + | n.s. | n.s. | n.s. |
| *Claims* | ∩ | ∩ | + | ∩ | ∩ |
| *Scope* | n.s. | n.s. | + | n.s. | ∩ |
| *Novelty* | ∩ | n.s. | ∩ | n.s. | ∩ |

Note: "n.s." indicates that the effect of the determinant on the patent influence was not significant ($p > 0.1$).

Findings also show that the breadth of technological base generally does not significantly affect the influence a patent exerts on subsequent technological developments. In other words, the analysis suggests that combining knowledge across either multiple or few domains does not significantly contribute to enhancing the impact of the resulting patented invention. This result may be explained by the research on knowledge search and recombination process according to which most innovative firms are those performing a balanced search process [34, 41]. In fact, on the one hand, excessive complexity [96] and lack of absorptive capacity [47], and, on the other hand, a limited set of technological opportunities to be combined [40] may, respectively, undermine search efforts towards the development of significant inventions. Nevertheless, a different result emerges when referring to a patent influence outside the biotechnology industry where a patent's technological breadth exerts a negative effect on the citations the patents receives from non-biotechnology patents. The costs of going along with complexity seem to outweigh the benefits of multiple recombinant possibilities, by significantly reducing the capability of organizations operating in different industrial domains to build on the technological solutions posed by such patented inventions [97].

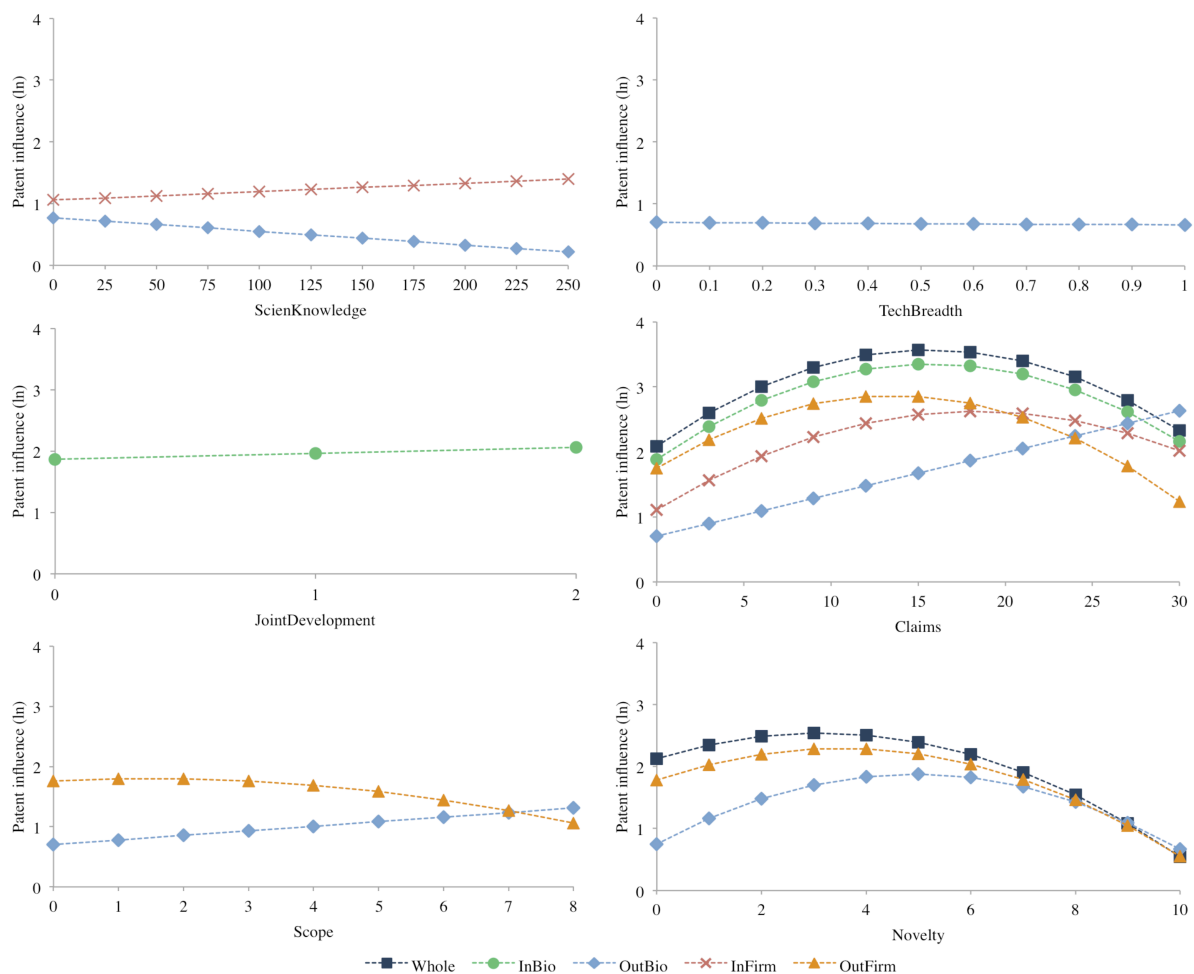

Figure 3. Determinants of patent influence (*ceteris paribus*, only significant relationships are reported, i.e. *p<0.1*).

The joint development of a patent presents a significant effect only when referring to the influence of the patent within the biotechnology domain. This effect is positive and may be closely related to the nature of the patent in our sample. The majority of joint patents (about 90%) occur between biotechnology firms or between biotechnology firms and large pharmaceutical companies. Being highly focused on specific strategic issues, these patents may result of interest especially to those firms operating within the biotechnology. This finding is in line with a recent study by Zidorn and Wagner [98], who found that alliances in the biotechnology industry are generally established to specialize in a certain research field, rather than to explore and enter into new markets and domains, hence making the influence of their innovative outcomes generally bounded within biotechnology.

The effect of the number of claims on the future citations a patent receives is curvilinear (inverted-U) in all the considered domains - except for the influence the patent exerts outside biotechnology, where the impact is linear and positive. Results suggest that as the number of original functionalities a patent claims increases the uncertainties associated with the domain of application of the patent also increase [99]. This may therefore reduce the probability that the patent is used as a valuable source for the subsequent technological developments.

The results on scope suggest the presence of a curvilinear effect on patent influence outside the assignee's organizational domain. The effect is however linear and positive on patent influence outside biotechnology. As expected, a patent of a wide industrial applicability is more likely to find successful applications outside the specific organizational and industrial domains, although in the former a trade-off emerges. This finding seems to suggest that a patent with a broader scope may be the result of specific strategic choices rather than innovativeness of the given patent *per se* [62]; thus, the patent may result in being technologically less valuable for other organizations.

Finally, the last determinant we investigated is the novelty of patents, that is, the extent to which the patent differs from the state of the art. Our analysis shows that its effect is curvilinear on patent influence as a whole and patent influence across the industrial and organizational boundaries. This suggests that a high degree of novelty may slow down the diffusion of the patented invention [73]. In fact, other firms, also operating outside the biotechnology industry, may encounter difficulties in absorbing and using the novel invention since their different and distant technological competences and knowledge bases. This may therefore hinder those firms' capability in innovating by following the new traced directions.

## 6. CONCLUDING REMARKS

In this paper, we put forward potential estimates of patent influence on subsequent technological developments: (i) the use of scientific knowledge, (ii) the breadth of the technological base, (iii) the existence of collaboration in patent development, (iv) the number of claims, (v) the scope of the patent, and (vi) the novelty. We measured patent influence as the number of citations received by subsequent patents. Citations were distinguished across four domains identified by the industrial and organizational boundaries. The results, based on an empirical analysis of 5671 biotechnology patents from 1976 to 2003, revealed that the contribution of the estimates to patent influence varies as the different domains of impact are considered.

Our study contributes to the research on patent estimation in three main ways. First, we conducted a multi-dimensional analysis of patent influence by differentiating the impact patents exerted within and outside its industrial domain and within and across the assignee's organizational boundaries. This shed more light on the meaning of patent influence on subsequent patented technologies, hence providing a better comprehension of how inventors and organizations may focus their resources on the different determinants according to the specific impact they aim at enhancing. Second, the present paper is one of the few attempts to analyze patent influence by presenting a rather broad investigation of the potential antecedents. In fact, we discussed the influence of a number of full-text patent indicators, thus showing how they differently affect the types of impact. Third, we provided an in-depth investigation of a highly complex sector, as the biotechnology industry, where patenting activity plays a fundamental role for understanding the dynamics that characterize the innovation processes.

The analysis offered several practical implications. In fact, by investigating patent influence on the basis of different perspectives, we encourage inventing individuals and firms to correctly identify the main patenting aim for a more effective resource allocation. We showed that patent influence differs across the industrial and organizational domains, thus highlighting that strategies and actions need to be carefully designed in the attempt to achieve the desired impact given contrasting effects of the considered determinants. For example, findings show that joint development activity makes patents technologically relevant within the biotechnology industry, while patents with a broader scope exert a stronger influence on the technological developments outside biotechnology. The evaluation of a firm's patent portfolio should therefore take into account these differences by weighting patent characteristics on the basis of which type of impact should be increased. An understanding of the different patenting results may also allow policy makers to define policies and aids to better address the required purposes. This is the case of the effect exerted by the use of scientific knowledge. A patent that is built more extensively on scientific knowledge represents an important base for subsequent technologies (inventions) of the patent's assignee. This provides evidence of the importance to create incentives for firms to strengthen their scientific base by, for example, the recruitment of scientists as well as the establishment of collaborations with research-oriented organizations.

It is worth noting the limitations of our analysis. First, despite the extensive and well-recognized role of forward citations in capturing the influence of patents [11, 12, 20], their use has been questioned. In particular, patent citations are often added by examiners rather than by assignees [11, 28], and examiners may lack the necessary resources for conducting an extensive search of the prior art [28] or decide to withhold citations for strategic reasons [29], hence creating some concerns about the reliability of citation-based indicators. Future research efforts should focus on the development of alternative proxies, based, for example, on multi-indicator approaches, to assess patent influence. Second, the paper did not investigate the potential interdependencies between the full-text patent indicators. Therefore, further studies could investigate the interacting effects between the determinants of patent influence in order to test the presence of complementary or substitutive effects. Third, the proposed research framework could be studied by introducing additional perspectives, such as strategic, geographical, and institutional. For instance, the geographical locations of patent assignee may contribute to explain differences across patent impact dimensions. Finally, the work analyzed only firms operating in the biotechnology industry and located in the US. Future studies may validate the results across different sectors and countries.

## ACKNOWLEDGMENTS

Daniele Rotolo acknowledges the support of the UK Economic and Social Research Council (award RES-360-25-0076 - "Mapping the Dynamics of Emerging Technologies"). The findings and observations contained in this paper are those of the authors and do not necessarily reflect the funder' view.## REFERENCES